# Toward a real synthesis of quantum and relativity theories: experimental evidence for absolute simultaneity

S. A. Emelyanov

Division of Solid State Electronics, Ioffe Institute, 194021 St. Petersburg, Russia

E-mail: sergey.emelyanov@mail.ioffe.ru

**Abstract**. We have demonstrated spatially-discontinuous quantum jumps of electrons at a distance as long as about 1cm. The effect occurs in a modified integer quantum Hall system consisted of a great number of extended Laughlin-Halperin-type states. Our observations directly contradict the no-aether Einstein's interpretation of special relativity together with the Minkowski's model of spacetime. However they are consistent with the aether-related Lorentz-Poincare's interpretation that allows absolute simultaneity. We thus strongly challenge the fundamental status of Lorentz invariance and hence break the basic argument against de Broglie-Bohm realistic quantum theory. We argue that both de Broglie-Bohm and Lorentz-Poincare theories are capable of providing a real synthesis of quantum and relativity theories. This synthesis is of such kind that quantum theory appears the most fundamental physical theory for which relativity is only a limiting case. In accordance with this hierarchy, quantum theory naturally resolves the problem of aether in Lorentz-Poincare's relativity. The role of aether could be played by a deeper Bohm-type undivided quantum pre-space, the relevance of which at any lengthscale directly follows from our observations.

> *It may be that a real synthesis of quantum and relativity theories requires not just technical developments but radical conceptual renewal.*
>
> *J. S. Bell (1987)*

## 1. Introduction

Today it is generally believed that there is no reasonable alternative to Einstein's interpretation of special relativity as well as to the Minkowski's model of spacetime, which prohibit absolute simultaneity even in principle. The Lorentz-Poincare's aether-based interpretation of relativity is rare in literature despite the fact that it does not contradict any experiments known up to now. This situation is partly due to the method of Einstein's original presentation of the special relativity in his seminal work of 1905 [1]. The method looks like an elegant, almost mathematical proof where macroscopic rod and clock are regarded as fundamental entities which need no any further clarification. However, the inconsistence of such method was stressed even by the author. In 1949, he wrote: *"The theory.... introduces two kinds of physical things, i.e., (1) measuring rods and clocks, (2) all other things, e.g., the electro-magnetic field, the material point, etc. This, in a certain sense, is inconsistent; strictly speaking measuring rods and clocks would have to be represented as solutions of the basic equations..., not, as it were, as theoretically self-sufficient entities..."* [2]. Nevertheless, this comment does not alter general belief in the ultimate truth of Einstein's theory.

However, the problem is that the interpretation of the special relativity is not "a matter of taste" but leads to significant physical consequences. One of them is the status of Lorentz transformations. Following Einstein's theory, Lorentz transformations reflect a fundamental symmetry, the so-called Lorentz invariance, which is relevant to the other physical theories, for instance, to quantum mechanics (QM). At the same time, even the founders of QM were aware of the fact that their theory is hardly consistent with the relativistic model of spacetime. On the Solvay Conference of 1927, Niels Bohr directly pointed out that "… *The whole foundation for causal*

*spacetime description is taken away by quantum theory, for it is based on [the] assumption of observations without interference…"* [3]. An even more formidable challenge is to reconcile relativity with the notion of instantaneous quantum jumping, which is an essential part of Schrödinger's wave mechanics. Schrödinger himself was so shocked by this notion that he was willing to give up the whole quantum theory. In this context, he addressed Bohr the following phrase: *"If all this damned quantum jumping were really here to stay, I should be sorry I ever got involved with quantum theory..."* Bohr's response is also well-known: *"… remember the Einstein derivation of Planck's [blackbody] radiation law. This derivation demands that the energy of the atom should assume discrete values and change discontinuously from time to time... You can't seriously be trying to cast doubt on the whole basis of quantum theory!"* [4]. So, on the one hand, the unprecedented success of QM in prediction of experimental observations left no doubt regarding its ultimate relevance but, on the other hand, it was hard to reconcile it with Einstein's relativity.

An extraordinary solution of the problem was proposed at the Solvay Conference by Bohr, Heisenberg and some other participants. It is known as the Copenhagen interpretation of QM, which actually is the standard QM known from textbooks. The key point of the solution is a radical denial of objectivity of the quantum world which appears to be real only insofar as some events, the so-called "measurements", take place in the macroscopic world. Bohr himself clarified this idea as follows: *"…There is no quantum world. There is only an abstract quantum mechanical description…"* [5]. Heisenberg spoke along the same line: *"… Reality is in the observation, not in the electron …"* [6]. In fact, it was an unprecedented case when realism, as a philosophical basis, was consciously rejected by a physical theory. In a sense, Bohr's solution was triggered by the fact that, at the time of Bohr, QM had a well-defined domain of applicability, that is, the microcosm. So, it seemed possible to separate subjective quantum world from objective classical world using some lengthscale criteria. At the same time, the advantage of Bohr's solution seems quite evident: henceforth any violation of Lorentz invariance in the quantum world cannot challenge a fundamental nature of this symmetry simply because this world is a subjective thing by definition.

However, the Copenhagen interpretation inevitably leads to serious conceptual problems. The most famous one is the so-called "measurement problem". In essence, the problem is that the formerly clear term "measurement" now acquires a truly mystical sound. Through the "measurement" some "measurer" literally creates, just like God, "something out of nothing" where something is a quantum object. Of course, a number of questions immediately arise in the context of the above "act of creation". One of them is who is the creator which could play the role of "measurer"? John Bell, for example, not without irony, tried to guess: *"…Was the world wave function waiting to jump for thousands of millions of years until a single-celled living creature appeared? Or did it have to wait a little longer for some more highly qualified measurer – with a Ph.D.?"* [7]. But that is not all. According to the well-known Bohr's complementarity principle, the "measurer" not only creates some object out of nothing but, simply by selecting the measuring devices, he (or she) literally "causes" the object to be either a corpuscle or a wave. As a brief comment to this principle, Bell put in: *"The justly immense prestige of Bohr has led to the mention of complementarity in most text books of quantum theory. One is tempted to suspect that the authors do not understand the Bohr philosophy sufficiently to find it helpful…"* [7].

Certainly, the above questions are not the only ones that appear in the context of the Copenhagen interpretation. However *"… the pioneers of quantum mechanics were not unaware of these questions, but quite rightly did not wait for agreed answers before developing the theory …"* [7]. But the point is the "agreed answers" haven't been found even up to now, i.e. almost hundred years later. As a result, a part of scientific community has come to the conclusion that we need no "agreed answers" at all simply because both mysticism and indeterminism are indeed the fundamental properties of microcosm. The other part of the community intuitively tries to avoid any mysticism reasonably believing that it strongly challenges one of the basic pillars of science – the principle of knowability. So, consciously or not, they follow a doctrine that could really appear only in the context of Bohr's QM. The doctrine sounds as "shut-up-and-calculate". This means that the formalism of Bohr's QM is indeed extremely successful but, at the same time, it is meaningless to think of what really lies behind it. Perhaps only a few persons refused to join any of the two camps. And one of them was Albert Einstein. On the one hand, he was convinced that "… *He [God] does not throw dice…*" [8]. But, on the other hand, he was no less convinced that "… *independence created by philosophical insight is … the mark of distinction between a mere artisan or specialist and a real seeker after truth …*" [9]. Unfortunately, the current trend is that the Einstein's decisive authority extends only over the theory of relativity but it breaks immediately when we are dealing with the quantum theory where his views are regarded as a stubborn conservatism.

However, there is an alternative realistic quantum theory though it is also rare in textbooks. This theory naturally avoids the conceptual problems of standard QM and its predictions are consistent with that of standard QM in all experimentally-assessable situations. It is based on Luis de Broglie's pilot-wave theory which actually is a typical "hidden-variable" theory. Historically, the pilot-wave theory was first presented on the Solvay Conference but was rejected in the view of the acceptance of Bohr's theory. Only in 50's it was revived and strongly developed by David Bohm and today it is known as de Broglie-Bohm pilot-wave theory or sometimes as Bohmian mechanics [10-12]. However, the current situation is that most physicists do not take seriously this theory. It is viewed at best as an apocryphal version of QM despite the fact that its comparison with standard QM has led Bell to the following conclusion: *"I think that conventional formulations of quantum theory, and of quantum field theory in particular, are unprofessionally vague and ambiguous. Professional theoretical physicists ought to be able to do better. Bohm has shown us a way…"* [7].

The reason for the disliking of pilot-wave theory is actually well-known. While the violation of Lorentz invariance in Bohr's theory is regarded to be more or less remissible, the same thing in the pilot-wave theory is regarded to be absolutely inadmissible. And the key point here is the status of wavefunction. In the pilot-wave theory, it is not an abstract description but rather a real physical entity. As it was especially stressed by Bell, *"… no one can understand this [Bohm] theory until he is willing to think of $\psi$ as a real objective field rather than just a "probability amplitude"* [7]. From the perspective of Einstein's relativity, this point is truly unacceptable. Indeed, let us take the familiar EPR experiment with entangled photons. It is seen immediately that, in terms of pilot-wave theory, EPR nonlocality acquires the characteristic features of a real physical effect, that is, a real action-at-a-distance. In this case, the impossibility of nonlocal signalling seems a peculiarity of EPR situation rather than a fundamental physical low.

Accordingly, the results of actual EPR experiments are currently viewed as a strong evidence for the ultimate truth of both Bohr's QM and Einstein's relativity [13-15]. On the one hand, they are fully consistent with the predictions of Bohr's QM but, on the other hand, EPR nonlocal correlations appear essentially non-causal, i.e. they cannot be used for any signalling. Moreover, the last point allows a purely mathematical proof known today as Eberhard's theorem [16]. However, in accordance with the common belief that quantum entanglement is the only possible physical reason for nonlocality, the Eberhard's theorem is usually interpreted in a wider sense, that is, as a mathematic proof of the impossibility of nonlocal signalling at all. The theorem thus appears to be quite fundamental so that it is often called the "no-communication theorem". And this status is clearly consistent with the fundamental status of Lorentz invariance.

However, the current relations between quantum and relativity theories are clouded. The point is they only coexist but not more. Almost centennial efforts to unify them have failed and it becomes more and more clear that unification is even inaccessible at least until both theories are in their generally-recognized forms. Today this point is declared to be one of the most fundamental physical challenges. However, a promising way to reach such unification was almost intuitively felt by John Bell who, just like Einstein, was unable to accept quantum mysticism. The idea was expressed in one of his interviews of 1986: *"Behind the apparent Lorentz invariance of the phenomena, there is a deeper level which is not Lorentz invariant ... what is not sufficiently emphasized in textbooks, in my opinion, is that the pre-Einstein position of Lorentz and Poincare, Larmor and Fitzgerald was perfectly coherent, and is not inconsistent with relativity theory ... I want to say there is a real causal sequence which is defined in the aether..."* [17]. So the idea is that we should rehabilitate the realistic quantum theory through the return to the aether-based Lorentz-Poincare relativity where Lorentz transformations are an emergent phenomenon.

In fact, taking a realistic view of the world, this status of Lorentz transformations seems quite natural. Indeed, they clearly are relevant only in the case of a movement that could be characterized by a certain speed. However, say, in the EPR situation, nothing could be characterized in such a way. It is interesting to note that, even in 1982, Karl Popper, who was rather a philosopher than physicist, made the following comment to the EPR experiment: *"… We have to give up Einstein's interpretation of special relativity and return to Lorentz's interpretation and with it to … absolute space and time… Whether or not an infinite velocity can be attained in the transmission of signals is irrelevant for this argument..."* [18]. However, his view is rather an exception. Most physicists still believe that any alternative to Einstein's relativity is impossible and this belief is so deep that even the most radical supporters of quantum realism still try to squeeze pilot-wave theory into the Procrustean bad of Lorentz invariance [19-21].

So the current situation is that the only thing that could really break a truly universal belief in Einstein's relativity is the observation of a physical phenomenon which would directly contradict this theory but, at the same time, would be consistent with the Lorentz-Poincare's theory. Actually, it would be only the observation of a purely quantum communication which currently is regarded to be fundamentally impossible. And in this work we report on such observation. The effect has become possible insofar as it is not related to quantum entanglement but rests on such a well-known thing as the spatially-discontinuous transport of quantum particles. So the novelty of our work is that we directly demonstrate that this transport may well be macroscopic and hence should be regarded as a peculiar quantum dynamics capable of providing absolute simultaneity.

2. **Beyond the no-communication theorem: spatially-discontinuous macroscopic quantum transport jumps of quantum particles**

One of the main "surprises" which is presented by quantum mechanics is a possibility for quantum particles to transit from one spatial region to another without staying in the intermediate regions. The simplest example of that kind is an electron transition between atomic orbits when not only energy changes discontinuously but also the spatial position of the electron. However in such form, in which this effect is known today, it seems an absurd to interpret it in terms of a peculiar quantum transport. The point is the characteristic lengthscale of these transitions is quite microscopic in all known cases and this fact is generally regarded as being of a fundamental character. At the same time, according to the ideas of standard QM, the microscopy of spatially-discontinuous transitions ultimately implies that we are dealing not with objective processes but rather with a method of description. So, it seems quite natural that no one has considered these transitions in the context of their compatibility with relativity principles.

However, the common belief in microscopy of spatially-discontinuous transitions is intuitive at best. The point is there are no fundamental limitations on their lengthscale in the formalism of standard QM. This fact allows us to propose the following *gedanken* experiment. Let us consider a one-dimensional electron orbit *C* which can be approximated by a circle with a macroscopic diameter *d*. Let there be also two local quantum levels *A* and *B* with lower energies, which are located as shown in Figure 1. Initially, only level *A* is occupied by electron while the other states are vacant. We expose the region *A* to light capable of providing quantum transition *A*→*C* while in the region *B* there is a number of local scatterers capable of providing quantum transition *C*→*B* during a characteristic time $\tau$. In terms of QM, this means that after the absorption of light quantum, the electron will appear on the level *B* during the time $\tau$.

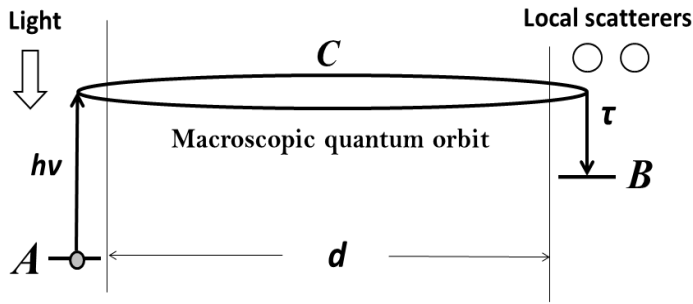

**Figure 1.** *Gedanken* experiment with a spatially-discontinuous quantum transport of electron between two local quantum levels (*A* and *B*) via a macroscopic-scale orbit-like state (*C*).

Actually, it is just the spatially-discontinuous transition from region *A* to region *B* avoiding any intermediate regions and we can talk about a real transport insofar as the distance *d* is macroscopic. But the most interesting point is that the characteristic time of the transport is independent of the distance *d* because it is determined by such local parameters as the concentration of scatterers in the region *B* and the efficiency of the scattering. Therefore, we potentially could fulfill the following relation: $d / \tau > c$. Thus, under the conditions of our *gedanken* experiment, the predictions of standard QM directly contradict Einstein's relativity as well as Minkowski's model of spacetime.

*2.1 Realisability of the 'gedanken' experiment*
Certainly, the most subtle point of our *gedanken* experiment is the existence of a macroscopic quantum orbit with an easy-to-control length. Indeed, at the Bohr-Einstein's times, such orbits were unknown so that our *gedanken* experiment could at best be regarded as a mind game. Today, however, it is not quite so. Recent development of nanotechnologies allows one to achieve such states of matter, in which the wavefunctions of quantum particles have an unusual shape and moreover their lengthscale may well be macroscopic. One of such states is the so-called integer quantum Hall (IQH) state [22-23].

As a rule, the IQH state appears in two-dimensional electronic systems at the temperatures below 10K and in a strong magnetic field higher than 1T. In the IQH state, there are two kinds of electron quantum states. Most electrons are in strongly localized microscopic orbits known as the cyclotron orbits. However, there is a relatively small number of quasi-one-dimensional orbits that are extended along the sample edges. This means that potentially the length of these orbits can be arbitrarily large. In the generally-recognized Laughlin-Halperin's theory, such orbits are often called the extended current-carrying states and just these orbits are responsible for the quantum Hall effect that emerges in presence of an electric bias [24-25]. As is well-known, for the discovering of this effect Klaus von Klitzing received the Nobel Prize in 1985 [26].

To clarify the origin of Laughlin-Halperin macroscopic edge states, let us address the energy diagram of an ideal IQH system (Figure 2a). Here the electrons are in equidistant energy levels, the so-called Landau levels, which are fully occupied below the Fermi level while otherwise they are empty. On Landau levels, the ordinary states are just the above-mentioned microscopic cyclotron orbits and the number of these orbits per level is roughly determined by their close packing within the sample area (Figure 2b). But the point is that near the sample edges there is a band bending that is an equivalent of the presence of a strong electric field pointed toward the sample center (see Figure 2a). As a result, close to the sample edges the electrons are in crossed electric and magnetic fields. In terms of a semi-classical model, this leads to an electron drift along the sample edges so that the direction of the drift is determined by the direction of external magnetic field. However, in terms of quantum mechanics, we should rather talk about quasi-one-dimensional current-carrying states with a characteristic width of the order of cyclotron radius. These are just the Laughlin-Halperin macroscopic orbit-like states.

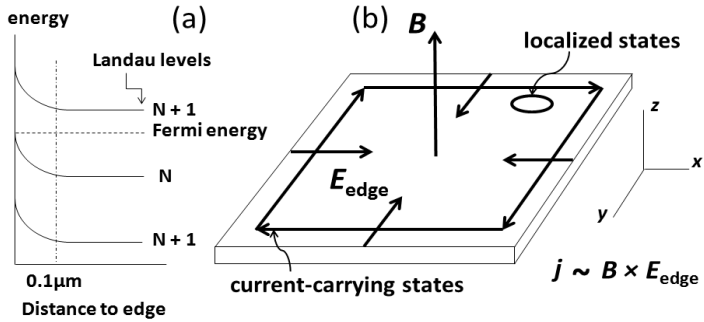

**Figure 2.** The origin of Laughlin-Halperin macroscopic-scale quasi-one-dimensional states. **(a)** Energy-band diagram of conventional IQH system close to sample edges. **(b)** Two types of quantum states in IQH system: localized states in the system interior and the Laughlin-Halperin states close to sample edges.

Generally speaking, the Laughlin-Halperin states seem suitable to implement our *gedanken* experiment. Actually, however, such implementation faces serious methodological difficulties. The point is that, in contrast to conventional electric measurements, in optical measurements we are always dealing with the effects related to the localized microscopic states. However, it turns out that the situation can be significantly improved if the external magnetic field has both quantizing and in-plane components while the entire two-dimensional layer is strongly asymmetric. The latter means that there is the so-called "built-in" transverse electric field within the layer. The field is due to the asymmetry of confining potential and may be as high as up to $10^5$V/cm [27]. One of the reasons for the asymmetry is the penetration of surface potential into the layer which may be very close to the sample surface as it is shown in Figure 3a [28]. In this case, the origin of the built-in field is similar to that of the edge field in a conventional IQH system. So we have got a system where not only edge electrons are in crossed electric and magnetic fields but the entire electron system (Figure 3b).

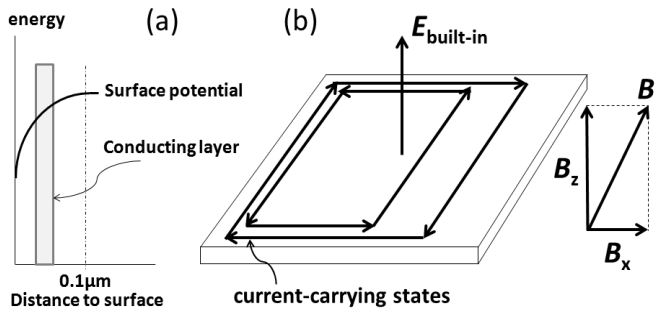

**Figure 3.** A method to provide Laughlin-Halperin-type states in the interior of an asymmetric IQH system. **(a)** Potential profile close to sample surface and typical position of a conducting two-dimensional layer. **(b)** Spatial distribution of Laughlin-Halperin-type states. Only two orbit-like states are shown with the opposite directions of spontaneous one-electron currents.

The solution of Schrödinger equation for a very simplified infinite model allows one to expect that, instead of the strongly localized cyclotron orbits, now we have got spatially-separated spontaneous currents throughout the entire conducting layer [29]. The currents are expected to be perpendicular to the in-plane component of magnetic field. Their origin is similar to that of Laughlin-Halperin spontaneous edging currents so that their characteristic width is also of the order of electron cyclotron radius. They should strongly differ from each other and even could have the opposite signs. In the case the above currents really exist in a finite system, they should be a mixture with the true Laughlin-Halperin edging states. So, their expected spatial distribution is shown in Fig. 3b though this picture certainly should be evidenced experimentally.

From this perspective, it is crucial that the spatial distribution of spontaneous current loops should not be exactly the same at different Landau levels. As a result, even in the absence of an electric bias, the resonant optical transitions between the Landau levels should give rise to local ohmic currents which could be detected. Moreover, these light-induced currents are easy to identify through their characteristic feature: they should be a strong non-periodic function of their position with respect to the sample borders.

Our real experiments strongly support the picture shown in Fig. 3b and their detailed description can be found in [30-31]. So, we are truly dealing with a great number of spatially-separated quasi-one-dimensional macroscopic orbits each of which is characterized by its own value of a one-electron spontaneous current. In a sense, the system is thus reminiscent of a gigantic *single* atom where the crystal lattice together with the valence electrons plays the role of a positively-charged nucleus while free electrons are distributed among atomic orbitals. Thus, in contrast to conventional solid state systems, these free electrons appear to be spatially-separated and therefore to be distinguishable.

*2.2 Demonstration of spatially-discontinuous macroscopic quantum transport*

Certainly, the system in Fig. 3b is much more suitable than a conventional IQH system to implement our *gedanken* experiment. With this system, one would propose the following version of the experiment in Fig. 1. Let us expose to light not the whole sample but only a part of it, which thus plays the role of the region *A* (see Fig. 1). Then, the role of the region *B* could play a distant region where the emergence of photo-excited electrons could be detected through the detection of a local ohmic current initiated by them. According to our guess, after the absorption of light quanta in the region *A*, photo-excited electrons could emerge in the region *B*, if their orbits cross this region and if they are lucky enough to interact with a local scatterer from this region. So, if a local current will truly be detected in the region *B*, then we will only have to prove that it can only be associated with a spatially-discontinuous transport of photo-excited electrons.

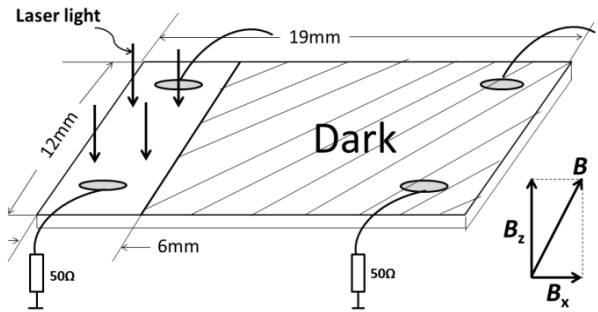

**Figure 4.** Realization of *gedanken* experiment in a modified IQH system with Laughlin-Halperin-type orbit-like states in the system interior. Only one third of the sample is exposed to light (region *A*). Signals are detected from two contact pairs: one pair is inside the laser spot while another one is about 1cm away from the spot so that the photo-excited electrons by no means can reach detectable area (region *B*) through continuous movement.

Our real experiment in accordance with the above scheme is shown in Fig. 4. We use single quantum well structures of InAs/AlGaSb type. The details of their structure as well as the experimental evidence for their asymmetry can be found in [31]. Typically, the width of the InAs conducting layer is 15 nm and the electron sheet concentration is about $10^{12}$cm$^{-2}$. The sample (19×12mm$^2$) has two pairs of short (1mm each) contacts which are located as shown in Figure 4. The sample is immersed in a superconducting magnet at the temperature as low as about 2K. We use the magnetic field of about 5T, which is tilted from the normal by the angle of 15° to provide the in-plane component. The sample is exposed to a spatially-uniform terahertz laser radiation. The source of radiation is optically-pumped pulsed ammonia laser with the intensity of incident radiation of up to 200W/cm$^2$ and with the pulse duration of about 40ns. The energy of light quanta is 13.7meV which roughly is the energy gap between the Landau levels under the experimental conditions so that resonant optical transitions to the lowest empty level have become possible. Electric signals from both contact pairs will be synchronously detected in a short-circuit regime when the resistance of load (50Ω) is about two order lower than that of conducting layer in the experiment.

Following our scheme, we will expose not the whole sample but only a one third of it, as shown in Fig. 4. So, one contact pair is inside the laser spot while another one is about 1cm away from the excited region. This means that the distance from the region *A* to the region *B* is *five orders higher* than the characteristic electrons' mean free path determined by local scatterers. In other words, the concentration of local scatterers is so high that any remarkable diffusion of photo-excited electrons is impossible at least during their characteristic lifetime on a higher Landau level, which is expected to be less than 0.1ns [32].

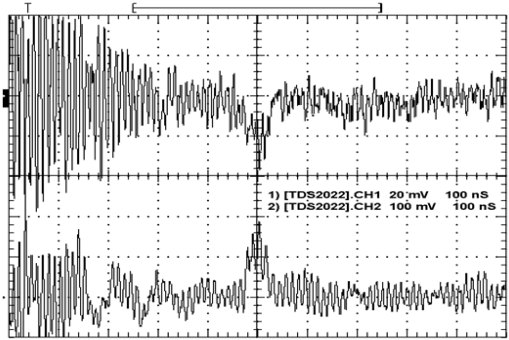

**Figure 5.** Typical tracks of synchronously-detection signals from two contact pairs in the experiment shown in Figure 4. Upper track – signal from the pair in the light (20mV/div.). Lower track – signal from the pair in the dark (100mV/div.). Timescale is 100ns/div.

Typical tracks from both contact pairs are shown in Figure 5 and, to be honest, the effect exceeds even our most optimistic expectations. Indeed, it is clearly seen that the signal in the dark is even *higher* than the signal in the light. This fact is consistent with the above ideas regarding a non-uniform distribution of currents throughout the sample and it clearly demonstrates that the current in the dark cannot be a "secondary effect" with respect to the current in the light. It is also clearly seen that the kinetics of both signals is so fast that any long-time diffusion or thermal effects are definitely irrelevant. Moreover, there is no delay between the signals at least with an accuracy of about 10ns. This means that to reach the region *B*, the photo-excited electrons should move toward this region with the speed of no less than $10^8$cm/s *without any collisions* at a distance of about 1cm. This is absolutely impossible because, as we mentioned above, their mean free path is about $10^{-5}$cm.

Finally, to avoid any loopholes, we repeat the experiment in Figure 4 but in a classic regime when the external magnetic field is fully in the well plane and therefore there is no Landau quantization. In this case, we also observe a signal from the pair in the light due to the so-called photo-voltaic effect [33]. However, despite the fact that the "classic" signal is two orders higher than the "quantum" signal from the pair in the light, no signal is observed from the pair in the dark.

Now it seems interesting to estimate the nominal "speed" of the transport from the region *A* to the region *B* as if we are truly dealing with a continuous movement of electrons. According to our ideas, the characteristic time of this transport is the so-called quantum relaxation time, that is, the time to destroy the coherence. In our structures it is about 0.3ps [27]. Thus, we immediately obtain the "speed" of about $3 \cdot 10^{12}$cm/s, i.e. two orders faster than the speed of light. Moreover, we can easily exceed this "speed", say, through the incorporation of additional scatterers into the region *B*.

## 3. Fundamental consequences: a new way to unify quantum and relativity theories

Our observations clearly have an applied aspect. But in this work we focus only on fundamental consequences which seem to have a greater importance.

One of these consequences is that the Einstein's relativity is experimentally-distinguishable from the Lorentz-Poincare's relativity and the experiment unambiguously supports the latter. However, this fact is crucial not only from the perspective of relativity but also from the perspective of quantum theory. The point is we break the basic argument against pilot-wave theory, that is, the lack of Lorentz invariance. We have thus got two quantum theories, standard QM and the pilot-wave theory, and both are equally successful in prediction of experimental observations.

However, their principle difference is that the former rests on mysticism and indeterminism while the latter rests on realism and determinism which is the basis of all current physical theories. As a result, the former encounters serious conceptual problems and can hardly be unified with relativity theory. Conversely, the latter naturally avoids conceptual problems and can easily be unified with relativity or, more precisely, with the Lorentz-Poincare's version of this theory. In this situation, the choice in favour of one of these theories seems self-evident though it clearly will take some courage to overcome the existing prejudices about quantum theory.

The important point is that the unification we mean is not a synthesis of equally fundamental theories, quantum mechanics and relativity, into a more general theory for which both ones are only approximations or limiting cases. Rather, we imply a hierarchy in the sense that quantum theory incorporates relativity as an approximation or a limiting case. Such a hierarchy allows one to overcome the natural difference between these theories on the level of their basic physical (but not philosophical!) concepts. This difference is related to the fact that relativity is inherently dealing with more or less autonomic objects. Otherwise, such notions as the "speed" or the "signalling" would appear meaningless. By contrast, quantum theory concerns essentially undivided systems for which any autonomy is a limiting case indeed.

Certainly, the indivisibility of quantum is the basis of both the standard QM and the pilot-wave theory. In a sense, Bohr's rejection of realism may be regarded as a sacrifice to reconcile quantum indivisibility with relativity. And this sacrifice seemed unavoidable at least at Bohr's times. This significant point was especially stressed by Bohm who wrote that "*… Bohr's rejection of hidden variables is … based on a very radical revision of the notion of what a physical theory is supposed to mean, a revision that in turn follows from the fundamental role which he assigns to the indivisibility of the quantum …*" [12]. Only now Bohr's sacrifice appears unnecessary. Moreover, instead of reconciliation, now we could rather talk about a harmony between quantum and relativity theories. The best manifestation of this harmony is the fact that quantum theory could help relativity to solve an important conceptual problem. It is the problem of physical meaning of aether.

However, to show the way of solving that problem we should first note the following thing. If the spatially-discontinuous jumps of quantum particles can indeed be macroscopic then we actually face a peculiar quantum dynamics. But any peculiar dynamics requires an adequate model of space which should be able to account for its characteristic features. As for the quantum dynamics, here the basic characteristic feature is the indivisibility of quantum. So, we need a model which would be adopted to account for this feature. But it turns out that such a model already exists and it is hardly surprising that the author of this model is David Bohm, i.e. one of the authors of the realistic quantum theory [11-12].

The basic idea of Bohm's model is that we give up such a seemingly eternal idea as the Cartesian order of space, which ultimately expresses the classical concept of spatially-separated interacting physical objects. Bohm himself characterized the "quantum" space (or pre-space) as follows: "*… pre-space would have a structure that was inherently conformable to the laws of the quantum theory. We would thus be free of the present incoherence of trying to force quantum laws into a framework of a Cartesian order that is really only suitable for classical mechanics…*" [12]. The "quantum" concept of spatial order was called the implicate order. Following Bohm's figurative comparison, this order differs from Cartesian order just like a holographic image differs from a mirror reflection. In the former case, each part of the image contains some information about the whole object and this information can potentially be unfolded. By contrast, in the latter case, each part of the reflection contains no information about the rest of the object. Leaving aside further details of Bohm's concept, we only would like to note that just this undivided "quantum" pre-space or undivided Universe could naturally play the role of aether or Bell's "deeper level" which is not Lorentz invariant.

Finally, it should be noted that Bohm did not actually suspect that the idea of quantum pre-space is so global. The point is he shared the common view, according to which quantum mechanics essentially concerns only microcosm. That is why he supposed that **"…** *processes that were associated with what we now call the Planck length would have to be described in what is an essentially quantum mechanical kind of space in which Cartesian notions of order would not even arise…"*. Moreover, he supposed that *"... as long as the quantum theory is valid, there is no way to demonstrate ... non-Lorentz invariance experimentally..."* [12]. However, our demonstration of macroscopic-scale non-Lorentz invariance clearly shows that quantum mechanics is not only and even not so much the theory relevant to microcosm. Rather, it is the deepest physical theory which is relevant at any lengthscale and therefore gives rise to a new – non-mechanistic – insight into the world.

**Acknowledgements**

The author is grateful to H.-T. Elze as well as to the other participants of the 6th International Workshop DICE2012 (Space-Time-Matter) for the friendly atmosphere during the Conference.